%% file: main.tex
\documentclass[entropy,article,submit,pdftex,moreauthors]{Definitions/mdpi}
\firstpage{1}
\pubvolume{1}
\issuenum{1}
\articlenumber{0}
\pubyear{2025}
\copyrightyear{2025}
\datereceived{ }
\daterevised{ } 
\dateaccepted{ }
\datepublished{ }
\hreflink{https://doi.org/} 

\let\P\relax
\DeclareMathOperator{\P}{P}
\DeclareMathOperator{\ID}{ID}

\DeclareMathOperator{\idiffe}{i^e_{diff}}
\DeclareMathOperator{\ispece}{i^e_{spec}}

\DeclareMathOperator{\idiffc}{i^c_{diff}}
\DeclareMathOperator{\ispecc}{i^c_{spec}}

\DeclareMathOperator{\idiff}{i_{diff}}
\DeclareMathOperator{\ispec}{i_{spec}}

\DeclareMathOperator{\idiffce}{i^{c/e}_{diff}}
\DeclareMathOperator{\ispecce}{i^{c/e}_{spec}}

\DeclareMathOperator{\ii}{ii}

\newcommand*{\argmax}{\operatornamewithlimits{argmax}\limits}

\usepackage{natbib, amsmath, amsthm, amssymb, amsfonts}

\input{macros}

\newlength{\maxwidth}
\newlength{\maxheight}
\Title{Intrinsic cause-effect power: the tradeoff between differentiation and specification}

\TitleCitation{Intrinsic cause-effect power: the tradeoff between differentiation and specification}

\Author{William G.P. Mayner$^{1,\ddagger}$\orcidA{}, William Marshall$^{2,\ddagger}$* and Giulio Tononi$^{1,}$*}

\AuthorNames{William G. P. Mayner, William Marshall and Giulio Tononi}

\isAPAStyle{%
       \AuthorCitation{Mayner, W.G.P., Marshall, W., \& Tononi, G.}
         }{%
        \isChicagoStyle{%
       \AuthorCitation{Mayner, W.G.P., Marshall, W., \& Tononi, G.}
        }{
       \AuthorCitation{Mayner, W.G.P., Marshall, W., \& Tononi, G.}
        }
}

\address{%
$^{1}$ \quad University of Wisconsin--Madison, USA, 53719 \\
$^{2}$ \quad Brock University, St Catharines, Canada, L2S 3A1}

\corres{Correspondence: gtononi@wisc.edu (G.T.); wmarshall@brocku.ca (W.M.)}
\secondnote{These authors contributed equally to this work.}

\abstract{
Integrated information theory (IIT) starts from the existence of consciousness and characterizes its essential properties: every experience is intrinsic, specific, unitary, definite, and structured. IIT then formulates existence and its essential properties operationally in terms of cause-effect power of a substrate of units. Here we address IIT's operational requirements for existence by considering that, to have cause-effect power, to have it intrinsically, and to have it specifically, substrate units in their actual state must both (\emph{i}) ensure the intrinsic availability of a repertoire of cause-effect states, and (\emph{ii}) increase the probability of a specific cause-effect state. We showed previously that requirement (\emph{ii}) can be assessed by the intrinsic difference of a state's probability from maximal differentiation. Here we show that requirement (\emph{i}) can be assessed by the intrinsic difference from maximal specification. These points and their consequences for integrated information are illustrated using simple systems of micro units. When applied to macro units and systems of macro units such as neural systems, a tradeoff between differentiation and specification is a necessary condition for intrinsic existence, \ie, for consciousness.
}

\begin{document}


\section{Introduction}

Integrated information theory (IIT,~\cite{albantakis2023}) attempts to explain consciousness by characterizing its essential phenomenal properties and then postulating that these phenomenal properties must be reflected operationally in properties of its physical substrate. IIT starts by noting that experience exists, and that every experience is intrinsic, specific, unitary, definite, and structured. IIT then translates these phenomenological axioms into physical postulates, where physical is understood operationally as taking and making a difference---\ie, as cause-effect power. Accordingly, IIT requires that a substrate exists (has cause-effect power) and that its cause-effect power is intrinsic, specific, irreducible, definite, and structured.

In IIT, the intrinsic, specific, unitary, and definite cause-effect power of a system is quantified by integrated information~\cite{albantakis2023, marshall2023system}. The definition of integrated information has been refined over time in an attempt to strengthen the mapping from essential phenomenal properties (axioms), to corresponding physical properties (postulates), and then to a mathematical framework for assessing the degree to which systems satisfy them~\cite{tononi2004, balduzzi2008, oizumi2014, albantakis2023, iitwiki, marshall2025}. Prior work introduced the requirement that units have a repertoire alternative states so that they can take and make a difference~\cite{balduzzi2008, oizumi2014}. The primary goal of the current work is to make explicit in the measure of system integrated information ($\varphi_s$) the requirement that a system must provide itself with a repertoire of alternative states.

For a system to have intrinsic cause--effect power, it must satisfy two complementary requirements. First, it must provide itself with a repertoire of alternative cause--effect states. Second, it must specify one of the potential cause-effect states by increasing the probability of one particular cause--effect state relative to the alternatives. In Section~\ref{sec:theory}, we operationalize these requirements through quantities grounded in IIT's intrinsic difference measure (ID)~\cite{barbosa2020}. We introduce a measure of \emph{intrinsic differentiation}, which is assessed as the difference between a system's conditional distribution and a maximally specific distribution, quantifying the degree to which alternative states are intrinsically available. \emph{Intrinsic specification} is assessed as the difference between the conditional distribution and maximal differentiation, quantifying the degree to which the system specifies a specific state~\cite{albantakis2023}. These quantities are incorporated into the mathematical framework, leading to an updated account of integrated information $\varphi_s$. Altogether, this framework clarifies the operational requirements for existence within IIT: a system with $\varphi_s > 0$ must both provide itself with a repertoire of alternatives and specify one of them, and it must do so as a unified whole. These dual requirements sharpen the definition of intrinsic information and integrated information, with important implications for identifying and characterizing substrates of consciousness.

In Section~\ref{sec:examples}, we illustrate the behavior of the refined measures in simple examples, beginning with the case of an isolated binary unit. We then extend the analysis to small systems of micro units and finally to macro units that represent coarse-grained substrates such as neural assemblies. The examples show how intrinsic differentiation grows with system size, how it interacts with intrinsic specification, and how their tradeoff provides a principled criterion for measuring intrinsic existence.

Finally, in Section~\ref{sec:discussion}, we discuss implications of this updated measure, especially as it relates to the fundamental role of intrinsic differentiation, which can be thought of as intrinsic `indeterminism'. We consider sources of indeterminism at the macro level---including fundamental stochasticity at the micro level, mappings from micro grains to macro grains, and fluctuations in background conditions impinging on the system---that may contribute to a system's intrinsic differentiation, and thus to its intrinsic cause--effect power.

\section{Theory}
\label{sec:theory}

A physical system can possess intrinsic cause--effect power only to the extent that it can provide itself with a repertoire of potential states. To illustrate this point, consider the case of a single-unit system where the unit's state transitions implement a deterministic COPY logic (\ie, if the current state is $s = 1$, both the past and future states are also $s = 1$ with probability 1). From the extrinsic perspective of an experimenter, we can set the system in state $s = 1$ and observe its behavior, then set the system into state $s = 0$ and observe its behavior again, repeat this many times, and finally conclude that it implements COPY logic. However, for the system itself, left to its own devices, there is only $s = 1$, with no possible alternatives. With no alternatives, the system cannot `make a difference' to itself, because from its intrinsic perspective, there is only the one option---there is no difference to be made.

This section formalizes the above intuition by (\emph{i}) revising the notion of intrinsic information from~\cite{albantakis2023} so that it now considers the extent to which a system provides itself with a repertoire of potential states, and (\emph{ii}) carrying this revision through to the definition of system integrated information. The following uses notation and conventions from~\cite{albantakis2023}, and further details about the formalism, including aspects not revised in the current work, can be found there.

\subsection{Mathematical preliminaries}\label{sec:prelim}

Throughout, we assume a discrete-time finite-state dynamical system governed by a transition-probability matrix (TPM) and adopt the causal-intervention semantics of the $\mathrm{do}(\cdot)$~operator~\cite{Pearl2009}. Let \(U=\{U_{1},\dots,U_{n}\}\) be a set of \(n\) binary units whose current state is $u\in\Omega_{U}$ with
$\Omega_{U} = \{0, 1\}^n$. The TPM for $U$ is

\begin{equation}
\label{eqn:tpm}
  \mathcal{T}_{U} \equiv p(\bar{u}\mid u)
  \;=\;
  \P\bigl(U = \bar{u} \mid \operatorname{do}(U = u)\bigr),
\end{equation}
where $u, \bar{u} \in \Omega_U$ are any two system states.

Generally, we are interested in subsystems $S\subseteq U$ in a current state $s \subset u \in \Omega_S$. Set operations on states, such as $s \subseteq u$, are imprecise expressions, as $s$ and $u$ are not technically sets, and this should instead be written $\{S = s\} \subseteq \{U = u\}$. However, hereafter we abuse notation and use the former expression, as it simplifies the exposition.

The complementary set $W = U\setminus S$ in state $w = u\setminus s$ is considered the background conditions for the purpose of evaluating the intrinsic cause-effect power of $S$. Before assessing intrinsic cause-effect power of $S$, the background conditions are accounted for by causal marginalization. At the micro grain, this process involves causally marginalizing the background units, conditional on the current state of $U$~\cite{albantakis2023}. The causal marginalization process in IIT differs slightly from the usual marginalization in probability theory. The marginalization is performed for each unit separately, and they are then recombined using a product, to ensure the resulting system TPM still has the conditional independence property. Letting $q_{c/e}(w \mid u)$ be the conditional distribution of $w$ given the current state $u$ (for causes, we are interested in the conditional distribution of the past state of $W$ and for effects the current state of $W$; see~\cite{albantakis2023}), the corresponding TPM is
\begin{equation}
    \label{eq:Tce}
    \mathcal{T}_{c/e} \equiv p_{c/e}(\bar{s} \mid s) = \prod_{i = 1}^{|S|} \sum_{\bar{w}} p(\bar{s}_i \mid s, \bar{w})q_{c/e}(\bar{w} \mid u), ~~ s, \bar{s}     \in \Omega_S, ~ u \in \Omega_U.
\end{equation}

When the elements of $S$ are macro units (grouped over multiple units and/or multiple updates), \eq\eqref{eq:Tce} is generalized by: (1) discounting micro connections extrinsic to $S$; (2) extending the modified micro TPM to sequences of updates; (3) causally marginalizing the background; and (4) compressing the resulting sequence probabilities into macro-state probabilities (see Section 2.2 of~\cite{marshall2025}).

The cause and effect TPMs ($\mathcal{T}_{c/e}$) form the basis of the intrinsic cause-effect power of $S$. Quantifying the cause-effect power of the system is done on using probability distributions extracted from the TPMs, and measuring the difference made to those distributions by some intervention. For example, the difference between intact and partitioned probability distributions is used to quantify integrated information. The $\ID$ is a unique measure of the difference between two probability distributions that satisfies a set of properties motivated by the postulates of IIT~\cite{barbosa2020}. Given two probability distributions, $p(s)$ and $q(s)$, the intrinsic difference is
\begin{equation}
    ID(p, q) = \max_s p(s)\log\left(\frac{p(s)}{q(s)}\right).
\end{equation}
It is worth noting that the $\ID$ is not a metric (it is asymmetric between $p$ and $q$, similar to the Kullback-Leibler divergence). Moreover, the $\ID$ can be decomposed into two terms: $p(s)$ is the selectivity, which describes how likely the state $s$ is to occur under $p$, while $\log p(s) / q(s)$ is the informativeness, which describes how much power $p$ has to bring about $s$ relative to $q$.


\subsection{Intrinsic differentiation}\label{sec:intrinsic-diff}

The novel measure we introduce here is \emph{intrinsic differentiation}, which quantifies the degree to which a system provides itself with a repertoire of potential states. The measure should be similar to entropy, in that it is zero when there is a perfectly deterministic system, and it should increase with decreasing determinism. Moreover, the units of intrinsic differentiation should be the same as those of intrinsic information and integrated information, \ie, \emph{ibits}~\cite{albantakis2023}). With these considerations in mind, we define the intrinsic effect differentiation for a current state $s$ and an effect state $s'$ as the $\ID$ between a maximally specific (deterministic) probability distribution for $s'$, and the conditional probability distribution the effect state given the current state,

\begin{equation}
  \idiffe(s, s') = \ID(p_{s'}(\bar{s}), p_e(\bar{s} \mid s)) = -\log(p_e(s' \mid s)),
\end{equation}
where
\begin{equation}
    p_{s'}(\bar{s}) = \begin{cases} 1 \text{ if } \bar{s} = s' \\ 0 \text{ otherwise.} \end{cases}
\end{equation}

The intrinsic cause differentiation for a current state $s$ and a cause state $s'$ is defined analogously, as the $\ID$ between a deterministic probability distribution for $s'$ and the conditional distribution of the cause state given the current state,
\begin{equation}
  \idiffc = \ID(p_{s'}(\bar{s}), p_c^\leftarrow(\bar{s} \mid s)) = -\log(p_c^\leftarrow(s' \mid s)),
\end{equation}
where $p_c^\leftarrow(\bar{s} \mid s)$ is the conditional probability of the cause state $\bar{s}$ given the current state $s$, and is computed from $p_c(s \mid \bar{s})$ using Bayes rule (see \eq\eqref{eq:bayes} below).

\subsection{Intrinsic information}\label{sec:intrinsic-info}

Earlier formulations quantified intrinsic information only by how much the current state specified a cause--effect state, ignoring whether the system provided itself with a repertoire of alternative states \cite{albantakis2023}. Perfectly deterministic systems could therefore achieve high intrinsic information despite having no intrinsically defined alternatives. The revised definition expands intrinsic information into two complementary components: intrinsic differentiation ($\idiff$), which quantifies how a system intrinsically defines its own alternatives, and intrinsic specification ($\ispec$), which quantifies how it intrinsically specifies one of those alternatives.

Noting that the intrinsic information defined in \cite{albantakis2023} only captures the latter of the two key aspects of intrinsic cause-effect power, here that quantity is renamed as the intrinsic specification. For a system $S$ in state $s \in \Omega_S$, the intrinsic specification of $s$ about an effect state $\bar{s}$ is defined as the $\ID$ between the conditional distribution of effect state given the current state ($p(\bar{s} \mid s)$) and the unconditional distribution of the effect state $p_e(\bar{s})$,
\begin{equation}
    \ispece(s, \bar{s}) = p_e(\bar{s} \mid s) \log \frac{p_e(\bar{s} \mid s)}{p_e(\bar{s})},
\end{equation}
where
\begin{equation}
    p_e(\bar{s}) = \frac{1}{|\Omega_S|}\sum_{\hat s \in \Omega_S} p_e(\bar{s} \mid \hat{s}).
\end{equation}

The intrinsic specification of $s$ about a cause state $\bar{s}$ is defined analogously, as a product of a selectivity term and an informativeness term,
\begin{equation}
    \ispecc(s, \bar{s}) = p_c^\leftarrow(\bar{s} \mid s) \log \frac{p_c(s \mid \bar{s})}{p_c(s)},
\end{equation}
where
\begin{equation}
    p_c(s) = \frac{1}{|\Omega_S|}\sum_{\hat s \in \Omega_S} p_c(s \mid \hat{s})
\end{equation}
and
\begin{equation}
    \label{eq:bayes}
    p_c^\leftarrow(\bar{s} \mid s) = \frac{p_c(s \mid \bar{s})}{\sum\limits_{\hat{s} \in \Omega_S} p_c(s \mid \hat{s})}.
\end{equation}
Here, the $p_c(s \mid \bar{s})$ term comes directly from $\mathcal{T}_c$; it is the conditional probability distribution of the current state of $S$ given a cause state $\bar{s}$, and $p_c(s)$ is the unconditional probability of the current state $s$. The additional term $p_c^\leftarrow(\bar{s} \mid s)$ plays the role of selectivity, quantifying the likelihood that the current state was preceded by the cause state.

Intrinsic specification quantifies how a system specifies its cause and effect. By the information postulate, the cause and effect should be specific. To specify a state, we appeal to the principle of maximal existence, which states: when it comes to a requirement for existence, what exists is what exists the most~\cite{albantakis2023}. In other words, the system specifies the cause and effect states that maximize its specific cause-effect power,
\begin{equation}
    s'_{c/e} = \argmax_{\bar{s} \in \Omega_S} \ispecce(s, \bar{s}), \qquad s \in \Omega_S.
\end{equation}

\begin{figure}[H]
\begin{adjustwidth}{-\extralength}{0in}
\includegraphics[width=\fulllength]{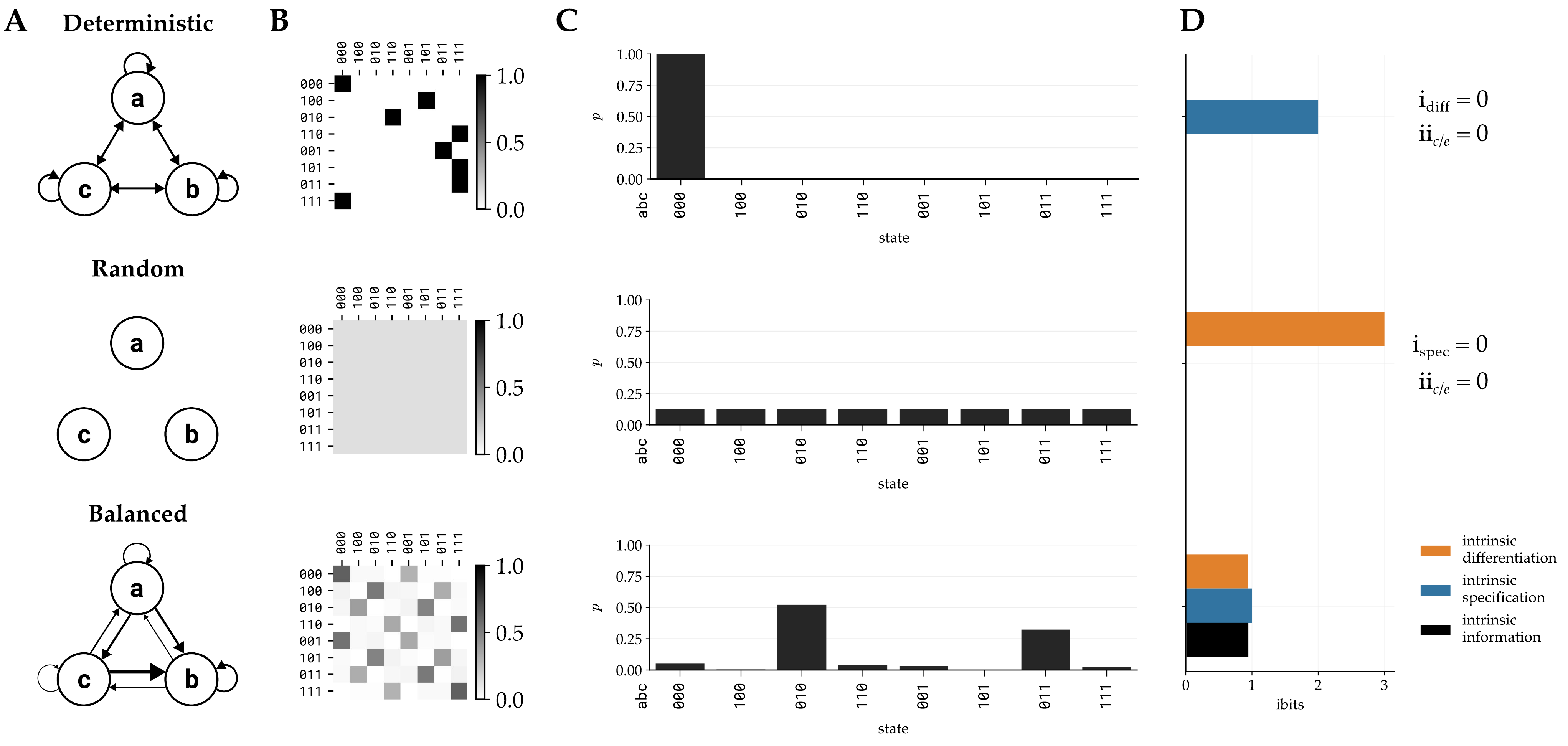}
\centering
\caption{
    \textbf{Intrinsic information captures the tradeoff between determinism and indeterminism.} (A) Causal diagrams of three systems with different dynamics: fully deterministic (top), uniformly random (middle), and a system balanced between those extremes (bottom). (B) The TPMs that describe each system. (C) Effect repertoires of each system for the state \texttt{000}. (D) Intrinsic differentiation $\idiff$, intrinsic specification $\ispec$, and intrinsic information $\ii(s)$ are shown for each repertoire, illustrating how intrinsic information captures the tradeoff between determinism and indeterminism. The fully deterministic system has $\idiff(s) = 0$, and thus $\ii(s) = 0$. The uniformly random system has $\ispec(s) = 0$ and so also has $\ii(s) = 0$. The balanced system, which defines several possible alternative effect states ($\idiff(s) > 0$) while also specifying one of them ($\ispec(s) > 0$), achieves positive intrinsic information $\ii(s) > 0$.
}
\label{fig:schematic}
\end{adjustwidth}
\end{figure}

Intrinsic differentiation and intrinsic specification capture the two requirements for cause-effect power that is intrinsic and specific: how a system intrinsically provides its own alternatives ($\idiff$), and how a system specifies one of those alternatives ($\ispec$). In IIT, the complementary principle to the principle of maximal existence is the principle of minimal existence, which states: when it comes to a requirement for existence, nothing exists more than the least it exists~\cite{albantakis2023}. Accordingly, we define the intrinsic cause and effect information as the minimum between intrinsic differentiation and intrinsic specification (Figure \ref{fig:schematic}),
\begin{equation*}
    \ii_{c/e}(s) = \min\{\idiffce(s), \ispecce(s)\}.
\end{equation*}

Likewise, both intrinsic cause and effect information are requirements for existence, and thus we define the overall intrinsic information of the system as the minimum between cause and effect,
\begin{equation}
    \ii(s) = \min\{\ii_c(s), \ii_e(s)\}.
\end{equation}

\subsection{Integrated Information}

A system that provides its own repertoire of potential states---and specifies one of those states---has intrinsic, specific, cause-effect power. Furthermore, if the system specifies its state as a unified whole, then it has intrinsic, specific, and irreducible cause-effect power. Integrated information quantifies how a system specifies its state as a whole, relative to how it specifies its state when cut into independent parts. The following is the same as the IIT 4.0 definition of $\varphi_s$ from~\cite{albantakis2023}, until \eq\eqref{eq:varphi}, when intrinsic differentiation is incorporated into the measure. 

For a single-unit system (a `monad'), the system is by definition a whole that is irreducible to parts. In this case, all the intrinsic information specified by the system is considered to be integrated information, and
\begin{equation}
    \varphi_s(s) = \ii(s).
\end{equation}

For systems with more than one unit, there are multiple ways to partition a system. For IIT, a directed partition is defined as a partition of the system such that each part has either its inputs cut, its outputs cut, or both. A directed partition has the form
\begin{equation}
    \theta = \{S_{\delta_1}^{(1)}, \ldots, S^{(k)}_{\delta_k}\} \in \Theta(S),
\end{equation}
where $S^{(1)}, \ldots, S^{(k)}$ is a partition of $S$ and each $\delta_i \in \{\leftarrow, \rightarrow, \leftrightarrow\}$, respectively indicating whether its inputs, outputs, or both are cut, and $\Theta(S)$ is the set of all directed partitions. For a given partition $\theta$ and a part $S^{(i)} \in \theta$, we can define the set of inputs $X^{(i)} \subset S$ to $S^{(i)}$ that are cut by $\theta$,
\begin{equation}
    X^{(i)} = \begin{cases} S \setminus S^{(i)} & \text{ if } \delta_i \in \{\leftarrow, \leftrightarrow\}, \\ \bigcup\limits_{\substack{j \neq i \\ \delta_j \in \{\rightarrow, \leftrightarrow\}}} S^{(j)} & \text{ if } \delta_i \in \{\rightarrow\},
    \end{cases}
\end{equation}
and the complementary set of inputs left intact, $Y^{(i)} = S \setminus X^{(i)}$.

For a given partition $\theta$, we can compute partitioned cause and effect TPM that describe the transition probabilities of the system after it has been partitioned, and cut connections are replaced with a uniform repertoire over alternative states,
\begin{equation}
    \mathcal{T}_{c/e}^\theta \equiv p_{c/e}^\theta(\bar{s} \mid s) = \prod_{j = 1}^n p^\theta_{c/e}(\bar{s}_j \mid s), \qquad s,\bar{s} \in \Omega_S,
\end{equation}
where $p_{c/e}^\theta(\bar{s}_j \mid s)$ is the partitioned distribution of unit $S_j \in S^{(i)}$,
\begin{equation}
    p_{c/e}^\theta(\bar{s}_j \mid s) = \|\Omega_{X^{(i)}}\|^{-1} \sum_{x^{(i)} \in \Omega_{X^{(i)}}} p_{c/e}(\bar{s}_j \mid x^{(i)}, y^{(i)}).
\end{equation}

The integrated effect information of a system $S = s$ over a partition $\theta \in \Theta(S)$ is the $\ID$ between the intact and partitioned repertoires, but evaluated specifically for the effect state identified by intrinsic specification $s'_e$:
\begin{equation}
    \varphi_e(s, \theta) = p_e(s'_e \mid s) \log\left(\frac{p_e(s'_e \mid s)}{p_e^\theta(s'_e \mid s)}\right).
\end{equation}
The integrated cause information over a partition $\theta \in \Theta$ is defined analogously, but again, using $p_c^\leftarrow$ for the selectivity term:
\begin{equation}
    \varphi_c(s, \theta) = p_c^\leftarrow(s'_c \mid s) \log\left(\frac{p_c(s \mid s'_c)}{p_c^\theta(s \mid s'_c)}\right).
\end{equation}

However, there are many ways to partition a system. Again, following the principle of minimal existence, we define the integrated information of the system as the integrated information over the system's minimum partition. The minimum partition is that which minimizes the integrated information after normalizing by the number of possible edges in the causal graph that span the parts (the maximum possible integrated information for a partition),
\begin{equation}
    \theta' = \argmax_{\theta \in \Theta} \frac{\varphi(s, \theta)}{\sum_{i = 1}^k |S^{(i)}||X^{(i)}|}
\end{equation}
and then
\begin{equation}
    \varphi(s) = \varphi(s, \theta').
\end{equation}

Moreover, both integrated cause information and integrated effect information are conditions for existence, as well as intrinsic differentiation and intrinsic specification. Again, by the principle of minimal existence, the integrated information of the system is the minimum among these quantities,
\begin{equation}
\label{eq:varphi}
    \varphi(s) = \min\{\varphi_c(s), \varphi_e(s), \ii(s) \}.
\end{equation}

Altogether, $\varphi_s(s)$ quantifies the intrinsic, specific, and irreducible cause-effect power of the system $S$ in state $s$. It accounts for three requirements of existence: that a system provides itself with a repertoire of alternative states (intrinsic differentiation), that it specifies one of its potential alternatives (intrinsic information), and that it does so as a whole, irreducible to independent parts (integrated information).

\section{Worked Examples}
\label{sec:examples}

In Section \ref{sec:theory}, we formalized the idea that intrinsic cause-effect power requires a system to provide itself with a repertoire of potential states. Intrinsic differentiation was introduced to quantify the extent to which a system provides itself with a repertoire of alternative states. The intrinsic differentiation measure is folded into the existing mathematical framework of IIT to define an updated version of system integrated information ($\varphi_s$) that accounts for a system's ability to provide itself with a repertoire of possible states.

In what follows, we present examples exploring how the updated account of system integrated information interacts with previously explored properties of the mathematical framework. Specifically, we consider (\emph{i}) the $\varphi_s$ value for a single-unit system (a `monad'); (\emph{ii}) how the update impacts the search for subsystems that maximize $\varphi_s$ (complexes); and (\emph{iii}) how the update influences whether cause-effect power peaks at macro grains. Calculations were performed with the PyPhi toolbox for IIT \cite{mayner2018pyphi}.

\subsection{Example 1: Monads}

A monad is a single-unit system. Monads, by definition, are integrated wholes that cannot be cut into parts. As such, all intrinsic information specified by a monad is considered to be integrated information. The intrinsic specification of a monad is maximized when the monad is deterministic (the past and future states of the monad are fully determined by the current state of the monad); however, in that case the intrinsic differentiation is zero. By contrast, the intrinsic differentiation of the monad is maximized when, given the current state of the monad, the potential past and future states are equally likely, but in this case the intrinsic specification is zero. Since intrinsic information (and thus also the integrated information of a monad) is the minimum between intrinsic differentiation and intrinsic specification, it follows that the $\varphi_s$ of the monad will be maximized for some intermediate level of determinism.

\begin{figure}[H]
\includegraphics[width=\textwidth]{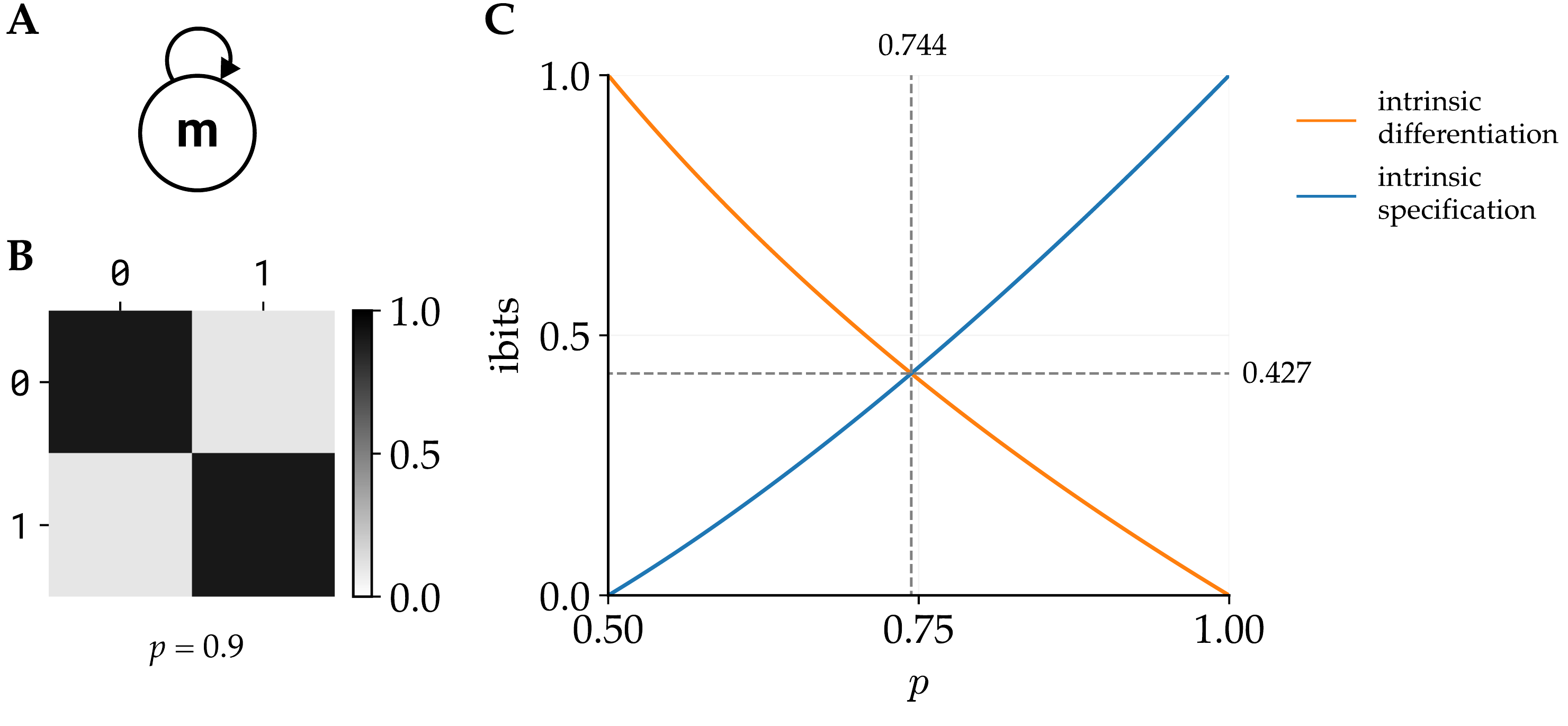}
\centering
\caption{\textbf{Example 1:} (A) a single-unit system (a monad). The unit stays in the current state with probability $p$, and switches to its other state with probability $1 - p$. (B) The transition probability matrix when $ p = 0.9$. (C) The intrinsic differentiation and intrinsic specification of the monad as a function of $p$. The intrinsic information (and thus integrated information) is maximized at $p = 0.744$, where $\varphi_s = 0.427$.}
\label{ex:monad}
\end{figure}

Consider the monad shown in Figure~\ref{ex:monad}A with corresponding transition probability matrix show in Figure~\ref{ex:monad}B. The function of the monad is an imperfect COPY gate: the monad remains in its current state with probability $p$, and switches states with probability $1 - p$, for some $p \in (0.5, 1]$ (the case where $p \in [0, 0.5)$ is symmetric, but in that case the monad would be an imperfect NOT gate). For this simple system, the cause and effect results are identical. For the system in the ON ($s = 1$) state (though the results are identical for the OFF ($s = 0$) state), the intrinsic differentiation of the system is
\begin{equation}
\idiffce(s, \bar{s}) = \begin{cases} -\log(p) & \text{ if } \bar{s} = 0 \\ -\log(1 - p) & \text{ if } \bar{s} = 1, \end{cases}
\end{equation}
and the intrinsic specification is
\begin{equation}
\ispecce(s, \bar{s}) = \begin{cases} p\log(2p) & \text{ if } \bar{s} = 0 \\ (1 - p)\log(2(1 - p)) & \text{ if } \bar{s} = 1. \end{cases}
\end{equation}
The specified cause and effect states for the system are
\begin{equation}
s'_{c/e} = \argmax_{\bar{s}} \ispecce(s, \bar{s}) = 0,
\end{equation}
Since $p > 0.5$. The resulting intrinsic information, and thus integrated information, is
\begin{equation}
\varphi_s(s) = \ii(s) = \min\left\{p\log(2p), -\log(p)\right\}
\end{equation}
The first term, $p\log(2p)$, is 0 at $p = 0.5$ and increasing on $p \in (0.5, 1]$, while the second term, $-\log(p)$, is 1 at $p = 0.5$ and decreasing on $p \in (0.5, 1)$. Thus, the maximum value of $\varphi_s$ occurs at the intersection of the two curves. Numerically solving for their intersection shows that the maximum value of $\varphi_s(s) = 0.427$ occurs at $p = 0.744$ (Figure~\ref{ex:monad}C).

\subsection{Example 2: Complexes}

A system with greater $\varphi_s$ than all overlapping systems is called a complex, and according to IIT such a system is a physical substrate of consciousness. It is therefore important to examine how the dual requirements for intrinsic differentiation and intrinsic specification affect the growth of $\varphi_s$ as system size increases. The previous example illustrated how the introduction of intrinsic differentiation results in a tension between indeterminism and determinism. The goal of this example is to understand how, if at all, this tension influences the potential for systems with a large number of units to be a maximum of $\varphi_s$.

\begin{figure}[H]
\begin{adjustwidth}{-\extralength}{0in}
\includegraphics[width=\fulllength]{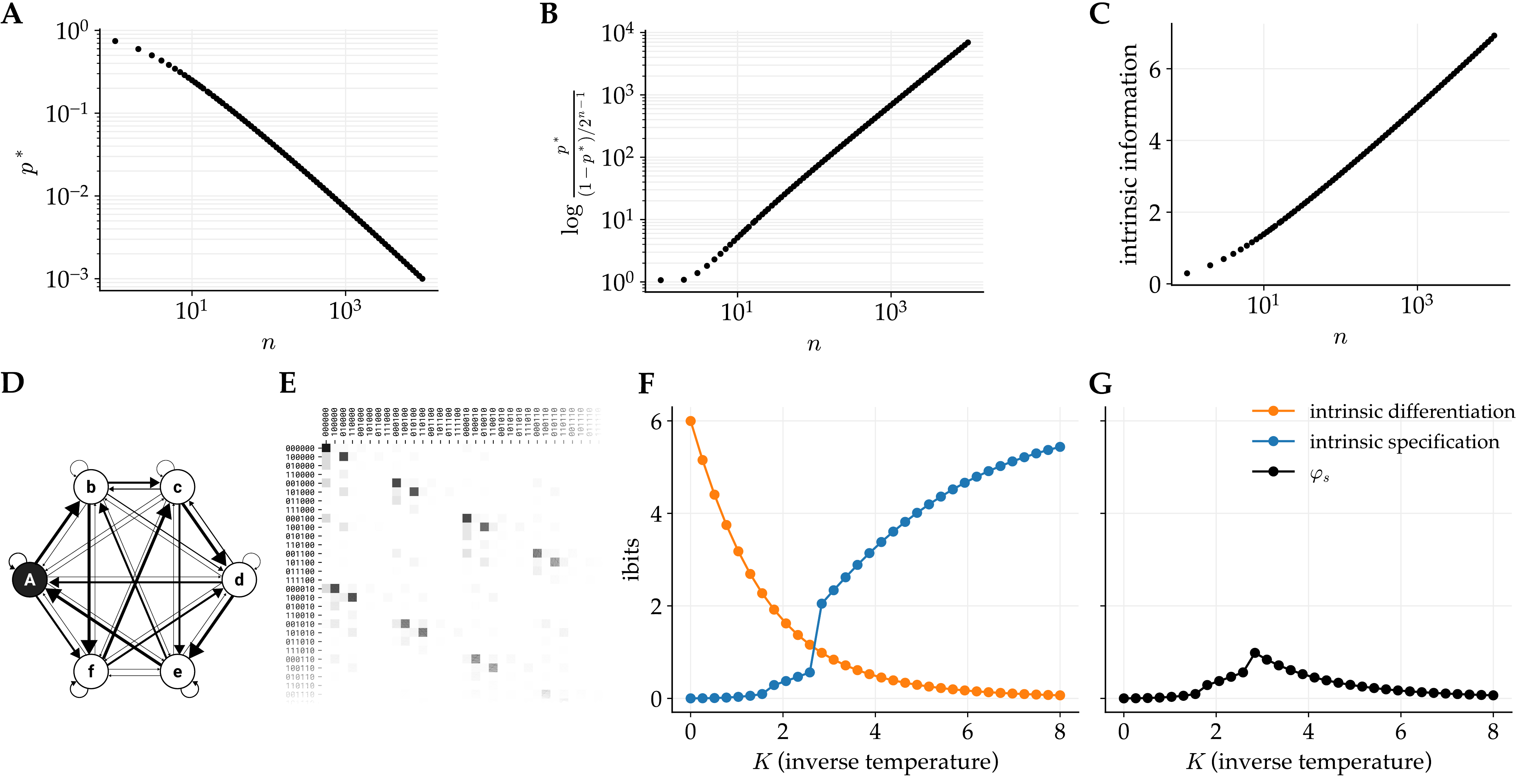}
\centering
\caption{\textbf{Example 2: Exploring the behavior of intrinsic information as a function of system size}. Consider a system of $n$ units that specifies a future state with probability $p$, with the remaining mass spread uniformly across potential alternative states. (A) The value $p$ that optimizes intrinsic information ($p^*$), as a function of system size. (B) The log ratio of $p^*$ to the probability of alternative states as a function of system size. (C) The value of intrinsic information at $p = p^*$ as a function system size. Taken together, we see that although $p^*$ decreases with system size, the ratio between it and the probability of alternative states is increasing, resulting in increasing intrinsic information as system size increases. (D) An example 6-unit system originally analyzed in \cite{albantakis2023}. The system has a temperature parameter $K$ that modulates the intrinsic differentiation of the system. (E) A partial transition probability matrix for the system, highlighting that the system specifies a particular state in each row. (F) As $K$ increases, the intrinsic differentiation decreases and the intrinsic specification increases. (G) Integrated information $\varphi_s$ is determined by intrinsic differentiation for low values of $K$, and intrinsic specification for high values of $K$. Overall, $\varphi_s$ is maximized at an intermediate value of $K$, balancing intrinsic differentiation and specification.}
\label{ex:complex}
\end{adjustwidth}
\end{figure}

Consider a system of $n$ units $S$ in a state $s$ such that its future state will be $s$ with probability $p$ and the remaining probability mass is spread uniformly across states, \ie for all $\bar{s} \in \Omega_S$ we have
\begin{equation}
	p_e(\bar{s} \mid s) = \begin{cases} p & \text { if } \bar{s} = s \\ (1 - p) /(2^n - 1) & \text { otherwise.} \end{cases}
\end{equation}
It follows that the unconstrained effect repertoire is uniform:
\begin{equation}
p_e(\bar{s}) = \frac{1}{2^n}, \quad \bar{s} \in \Omega_S.
\end{equation}
For such a system, the specified effect state is $s'_e = s$, and the corresponding intrinsic differentiation and specification are
\begin{equation}
\label{ex2:diff}
	\idiffe(s, s) = -\log(p)
\end{equation}
and
\begin{equation}
\label{ex2:spec}
	\ispece(s, s) = p\log(p2^n).
\end{equation}
The value of $p$ that maximizes the intrinsic effect information ($p^*$, the intersection of \eq\eqref{ex2:spec} and \eq\eqref{ex2:diff}) decreases as the system size increases (Figure~\ref{ex:complex}A), resulting in increased intrinsic differentiation. Additionally, the log ratio between $p^*$ and the $1 / (2n)$ increases faster than $p^*$ decreases (Figure~\ref{ex:complex}B), resulting in increased intrinsic specification. Finally, the ratio between $p^*$ and $(1 - p) / (2^n - 1)$ is increasing with system size (Figure~\ref{ex:complex}C), despite the decrease in $p^*$ in absolute terms. This indicates that, for this class of systems, the optimal behavior for maximizing intrinsic information in the limit of large $n$ is for the system to specify one effect state with probability $p$ greater than that of all other potential states.

To further explore the above theoretical argument, we consider an example system of $n = 6$ units (Figure~\ref{ex:complex}D) that was originally presented in \cite{albantakis2023} (Figure~6D in the referenced paper). The example includes a determinism (temperature) parameter $K = 4$ which results in the full 6-unit system being identified as the complex. In the current work, we analyzed this system using the updated account of $\varphi_s$ and examined the behavior of the system as $K$ is varied. The size of the largest complex varies with $K$, with the full 6-unit system being a complex for $K \gtrapprox 0.775$ and the system breaking down into two-unit complexes for $K \lessapprox 0.775$. It is worth noting that this is not due to the introduction of intrinsic differentiation, as the intrinsic differentiation only affects $\varphi_s$ when $K \gtrapprox 2.839$ (Figure~\ref{ex:complex}F).

\subsection{Example 3: Intrinsic Units}

The final example concerns IIT's requirement that a complex is the set of units that maximizes $\varphi_s$ across all subsets, and across all grains. The framework for assessing cause-effect power across grains is described in \cite{marshall2025}. A key aspect of the framework is that potential macro units must satisfy the postulates by being `maximally irreducible within', which essentially means that when treated as a subsystem they have greater $\varphi_s$ than any potential subsystems that can be defined from within (but not necessarily any supersets or partially overlapping sets). In this example, we revisit the question of whether cause-effect power can peak at macro grains with the updated account of $\varphi_s$, and whether intrinsic differentiation plays a role in determining if a unit satisfies the maximally irreducible within condition, or if a system of macro units has greater $\varphi_s$ than the corresponding micro units.

\begin{figure}[H]
\includegraphics[width=\textwidth]{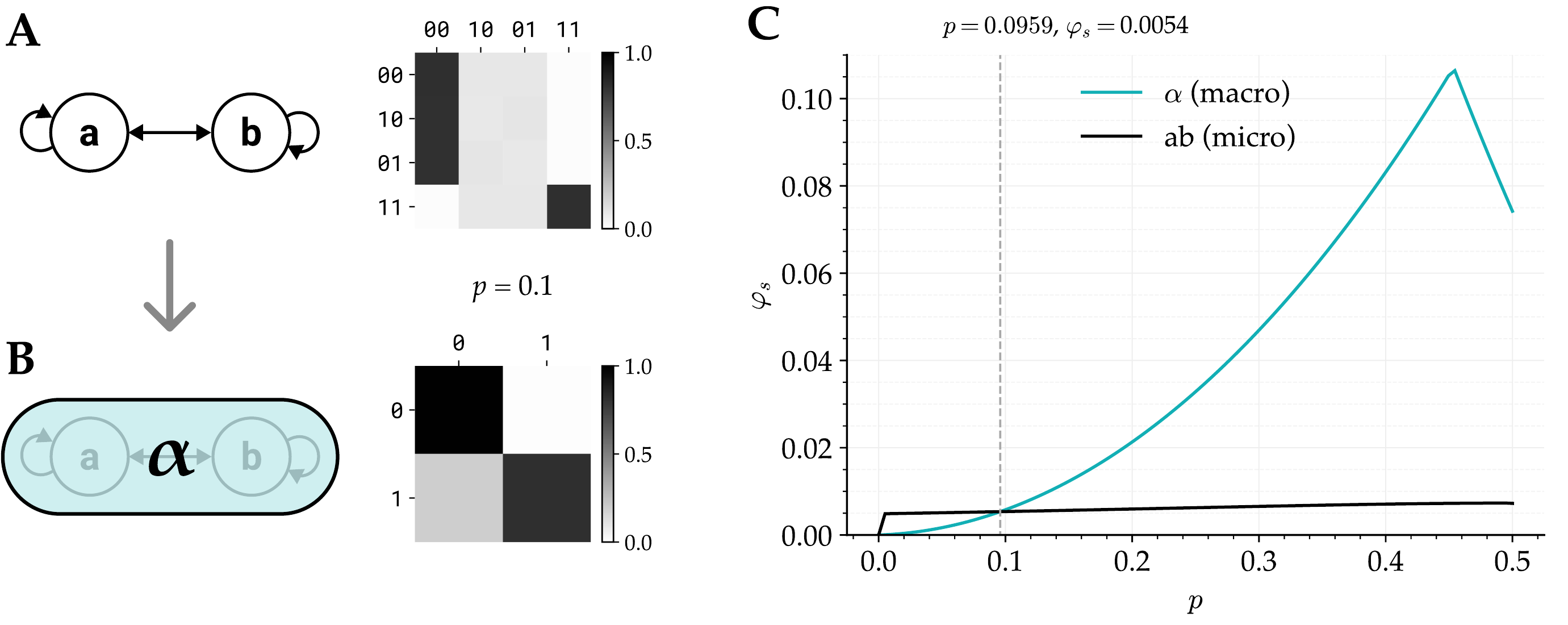}
\centering
\caption{\textbf{Example 3: Intrinsic units.} (A) A two-unit micro system implementing an imperfect AND logic with parameter $p$. The transition probability matrix is shown for $p = 0.1$. (B) The system $S = \{a, b\}$ satisfies the `maximally irreducible within' condition of \cite{marshall2025}, and can thus be considered as a macro unit. The macro unit $\alpha$ is defined by mapping states $(a, b) = (0, 0), (0, 1), \text{ and } (1, 0)$ to $\alpha = 0$, and $(a, b) = (1, 1)$ to $\alpha = 1$. The resulting transition probability matrix for $\alpha$ is shown. (C) $\varphi_s$ of (A) and (B) for different values of $p \in (0, 0.5)$. For $p < 0.096$, we find $\varphi_s(\{a, b\}) > \varphi_s(\alpha)$, while for $p > 0.096$, $\varphi_s(\alpha) > \varphi_s(\{a, b\})$. Thus, whether the macro or micro system has higher integrated information depends on intrinsic differentiation (via $p$).}
\label{ex:iu}
\end{figure}

For this example, we recreate a minimal example from \cite{marshall2025} (Figure~4 in the cited paper). The example starts with a two-unit system of micro units $S = \{A, B\}$ that each approximate the function of an imperfect AND gate (Figure~\ref{ex:iu}A). When both $A$ and $B$ are in the OFF state $s = (0, 0)$, then they will be ON in the future state with probability $p$. When $A$ is ON but $B$ is OFF, the probability that $A$'s future state is ON remains $p$, but the probability that $B$'s future state is ON is marginally increased to $p + 0.01$ (and vice versa when $A$ is OFF but $B$ is ON). When both $A$ and $B$ are ON, their future states will be ON with probability $1 - p$.

In \cite{marshall2025}, the example is presented using $p = 0.05$. In this case, $\alpha = \{A, B\}$ (Figure~\ref{ex:iu}B) satisfies the requirement of maximally irreducible within. When considered as a macro system $S = \{\alpha\}$, it has greater cause-effect power than the corresponding micro system $S = \{A, B\}$. In the current work, we reanalyze the system with the inclusion of intrinsic differentiation and compute $\varphi_s$ for $p \in (0, 0.5)$. The potential macro unit $\alpha = \{A, B\}$ satisfies the maximally irreducible within criteria for all $p \in (0, 0.5)$, and, moreover, the macro monad has greater $\varphi_s$ than the corresponding micro system for $p \in (0.096, 0.5)$ (Figure~\ref{ex:iu}C).

Thus, with the introduction of intrinsic differentiation, it is still possible to define intrinsic macro units, and those units may have greater cause-effect power than the corresponding micro systems. However, the outcome depends on the level of determinism in the system, and it may be the case that the intrinsic differentiation at the macro level prevents the macro system from outperforming the micro system.

\section{Discussion}\label{sec:discussion}

In this work, we refined the operational account of intrinsic cause--effect power within IIT by introducing the complementary concepts of intrinsic differentiation and intrinsic specification. We argued that a system must both provide itself with a repertoire of alternatives and specify one among them in order to have intrinsic existence. Building on these requirements, we defined intrinsic information as the minimum of differentiation and specification, and incorporated this measure into the calculation of $\varphi_s$. Through worked examples ranging from single-unit systems (monads) to larger complexes and macro-level units, we illustrated how the revised framework clarifies the role of indeterminism in IIT, and its impact on identifying complexes. Together, these results present a more complete account of the conditions under which systems can be said to possess intrinsic cause--effect power.

The central motivation for this refinement is conceptual: to ensure that the mathematical formulation of intrinsic cause--effect power aligns with IIT's postulates. Among these, the postulate of intrinsicality is primary: existence (intrinsic cause-effect power) must be defined from the perspective of the system itself, not relative to an outside observer. By explicitly requiring both differentiation (the system provides itself with a repertoire of alternative states) and specification (the system specifies one among those alternatives), the revised framework secures a closer link between the formal measures and the axioms they are intended to capture. In this way, intrinsic information captures precisely the two requirements that a system must satisfy to exist intrinsically.

A key consequence of introducing intrinsic differentiation is that it renders explicit a tradeoff between determinism and indeterminism. Intrinsic differentiation essentially captures indeterminism in the system, however without the standard interpretation: rather than something extra (`noise') added to the system, the indeterminism is intrinsic to the system and a requirement for its existence. Purely deterministic systems provide no genuine alternatives, and thus their intrinsic differentiation is zero, while purely random systems specify no state, leaving intrinsic specification at zero. Only systems that balance these two extremes can have non-trivial intrinsic information. This principle was clearly illustrated in the monad example, where integrated information peaked at an intermediate $p$, and it carried through to larger systems: as system size increased, differentiation naturally grew but had to be contrasted with sufficient specification. Similarly, in the analysis of macro systems, the extent to which coarse-grained units outperformed their micro constituents depended on whether an appropriate balance between determinism and indeterminism was maintained.

The requirement of intrinsic differentiation holds regardless of the spatiotemporal grain of the system. Quantum mechanics provides a well-established paradigm for micro-grain differentiation, most clearly in the context of collapse models that posit a wave function supporting potential alternative states~\cite{Bassi2023}. The interpretation of these models, and the precise ontological status of quantum indeterminism, remains subject to debate, with some accounts aligned with the current work (\eg, promoting indeterminism as an intrinsic feature of physical reality~\cite{delsanto2023}).

At the macro grain, differentiation does not simply disappear once it is present at the micro grain, but is instead reshaped by the process of mapping from micro to macro. Some of this differentiation is a direct extension of the differentiation of the underlying micro grain, inherited when micro states are grouped into a macro state. Yet, because cause--effect power is assessed intrinsic to the macro grain, additional sources of differentiation arise. One is the percolation of background conditions across a macro updates, where fluctuations external to the macro unit mediate cause-effect power. Another stems from uncertainty about the precise initial configuration and dynamic evolution of micro units constituting a macro unit, since many micro patterns may correspond to the same macro outcome. Together, these factors mean that macro-level systems often display more intrinsic differentiation than their micro counterparts.

The requirement of some degree of differentiation (\ie, indeterminism) also resonates with longstanding ideas about criticality and metastability in brain dynamics~\cite{wilting2019}. Neural systems operate near the edge of chaos, where activity is neither frozen into rigid determinism nor dissolved into unstructured randomness~\cite{beggs2003}. The present framework emphasizes that differentiation is not a nuisance to be minimized, but a necessary ingredient for intrinsic existence. A balance between differentiation and specification (\ie, indeterminism and determinism) provides the conditions under which neural substrates can maximize their intrinsic cause--effect power, revealing a further alignment of the principles of IIT with accounts of criticality in complex biological systems~\cite{mediano2022}.

\section*{Acknowledgements}

We thank Larissa Albantakis and Graham Findlay for their valuable comments on the manuscript.

\section*{References}

\bibstyle{plos2015}
\bibliography{references.bib}

\end{document}

%% file: macros.tex
\usepackage{xspace}

\newlist{panels}{enumerate*}{3}
\setlist[panels]{label=\textbf{(\Alph*)}}

\newcommand*{\eq}{Eq.~}

\newcommand*{\ie}{\emph{i.e.}\xspace}
\newcommand*{\eg}{\emph{e.g.}\xspace}

\renewcommand{\phi}{\varphi}

\renewcommand{\Phi}{\mathrm{\emph{\textPhi}}}